\documentclass[iop]{emulateapj}

\usepackage{graphicx}
\usepackage{amsmath}
\usepackage{subfigure}
\usepackage{float}
\usepackage{natbib}
\usepackage{hyperref}

\begin{document}

\title{Laser-only adaptive optics achieves significant image quality gains compared to seeing-limited observations over the entire sky}

\author{Ward S. Howard\altaffilmark{1}, Nicholas M. Law\altaffilmark{1}, Carl A. Ziegler\altaffilmark{1}, Christoph Baranec\altaffilmark{2}, Reed Riddle\altaffilmark{3}}
\altaffiltext{1}{Department of Physics and Astronomy, University of North Carolina at Chapel Hill, Chapel Hill, NC 27599-3255, USA}
\altaffiltext{2}{Institute for Astronomy, University of Hawai'i at M$\mathrm{\bar{a}}$noa, Hilo, HI 96720-2700, USA}
\altaffiltext{3}{Division of Physics, Mathematics, and Astronomy, California Institute of Technology, Pasadena, CA, 91125, USA}
\email[$\star$~E-mail:~]{wshoward@unc.edu}

\begin{abstract}
Adaptive optics laser guide star systems perform atmospheric correction of stellar wavefronts in two parts: stellar tip-tilt and high-spatial-order laser-correction. The requirement of a sufficiently bright guide star in the field-of-view to correct tip-tilt limits sky coverage. Here we show an improvement to effective seeing without the need for nearby bright stars, enabling full sky coverage by performing only laser-assisted wavefront correction. We used Robo-AO, the first robotic AO system, to comprehensively demonstrate this laser-only correction. We analyze observations from four years of efficient robotic operation covering 15,000 targets and 42,000 observations, each realizing different seeing conditions. Using an autoguider (or a post-processing software equivalent) and the laser to improve effective seeing independent of the brightness of a target, Robo-AO observations show a 39$\pm$19\% improvement to effective FWHM, without any tip-tilt correction. We also demonstrate that 50\% encircled-energy performance without tip-tilt correction remains comparable to diffraction-limited, standard Robo-AO performance. Faint-target science programs primarily limited by 50\% encircled-energy (e.g. those employing integral field spectrographs placed behind the AO system) may see significant benefits to sky coverage from employing laser-only AO.
\end{abstract}
\keywords{instrumentation: adaptive optics --- instrumentation: tip-tilt --- instrumentation: guide star --- instrumentation: sky coverage --- techniques: GenSTAC --- exoplanets: Kepler Objects of Interest}

\maketitle

\section{Introduction}
Correcting atmospheric distortion of stellar wavefronts involves two components: tip-tilt (e.g. stellar image displacement) and point-spread-function (PSF) irregularities. The image quality of laser adaptive optics suffers without a sufficiently-bright guide star nearby to correct tip-tilt error \citep{1992A&A...261..677R}. High-resolution-imaging science programs that include faint targets are thus susceptible to tip-tilt errors. 
\par In the case of \textit{Kepler} Objects of Interest (KOIs -- i.e. stars that host planet candidates), follow-up using high-resolution observations is a key step to rule out false-positive scenarios. \textit{Kepler} looks for the dip in brightness due to transits of the planet in front of the host star, and uses the transit depth and stellar radius to estimate planetary radius \citep{2010ApJ...713L..79K, 2006ApJ...644.1237O, 0004-637X-791-1-35}. Any blending of associated stars on the sky to the host dilute the transit depth and artificially decrease the calculated planetary radius \citep{0004-637X-730-2-79}. \citet{2016arXiv160503584Z} notes that faint hosts have received less focus in high-resolution follow-up efforts. 
\par For example, M-dwarfs, the most populous stellar type in the Galaxy \citep{2003PASP..115..763C}, are a class of intrinsically faint stars; Earth-sized planets orbit within the habitable zones of approximately one in six M-dwarfs \citep{DressingCharbonneau}. Although M-dwarfs account for only 2\% of all KOIs, the percentage of M-dwarfs amongst the faintest 12\% of KOIs is three times higher than in the overall KOI population$^1$.\footnotetext[1]{https://exoplanetarchive.ipac.caltech.edu/} The ability to detect companion stars amongst the faintest KOIs disproportionately affects the confirmation of M-dwarf rocky planets. 
\par A KOI companion-star survey with Robo-AO, the first autonomous laser guide star adaptive optics system \citep{2014ApJ...790L...8B}, has already observed 3857 KOIs due to its low observation overheads \citep{0004-637X-791-1-35,2016AJ....152...18B,2016arXiv160503584Z}. With hundreds of KOIs too faint for full Robo-AO post-facto image registration ($m_V > 15.5$), even modest gains in resolution above the seeing-limit allow discovery of companions deep within the \textit{Kepler} $\sim$4\arcsec pixel scale \citep{2041-8205-713-2-L115}.
\par In order to increase AO coverage of faint targets, methods to minimize tip-tilt error have been developed. The standard approaches are natural guide star (NGS) AO, laser guide star (LGS) AO \citep{1985A&A...152L..29F}, off-axis tip-tilt correction, e.g. \citep{Steinbring, 2013aoel.confE..51T,Wizinowich}, and laser-only AO \citep{2008Msngr.131....7D}. NGS AO employs bright guide stars to correct both tip-tilt and high-order errors. LGS AO supplies a laser guide star to correct high-order errors, although a natural guide star is still necessary to correct tip-tilt. Off-axis tip-tilt correction employs an off-axis guide-star with respect to the observation target for approximate tip-tilt correction. Laser-only AO employs an LGS AO system, but with no natural guide-star, and hence no tip-tilt correction.  
\par Correcting high-order errors requires a brighter guide star than does correcting tip-tilt, vastly expanding the sky coverage of LGS AO over NGS AO. Even with LGS, the necessity of sufficiently-bright guide stars significantly limits AO sky coverage \citep{1992A&A...261..677R}. 
\par For large apertures, off-axis tip-tilt error is reduced compared to smaller 1- or 2-meter aperture AO systems \citep{Hardy1998}. Off-axis guide stars approximate the tip-tilt of the target star, improving effective seeing. Using the Altair AO system on the Gemini-North telescope \citep{1998SPIE.3353..611R}, \citet{2013aoel.confE..51T} designed and tested LGS + Peripheral WFS 1 (LGS+P1 henceforth), an off-axis guiding mode with low-Strehl ``super-seeing." LGS+P1 NIR observations of $10^1$ to $10^2$ targets give 2-to-3x improvement in effective seeing \citep{2013aoel.confE..51T}. However, off-axis guiding is intended for large apertures, as scintillation and tip-tilt anisoplanatism decrease as aperture increases. 
\par Also relying upon reduced tip-tilt for large apertures, \citet{2008Msngr.131....7D} and references therin explore the use of an LGS system while foregoing the tip-tilt guide star entirely. Employing the VLT in K-band for a handful of observations, they demonstrate significant improvement to FWHM and encircled energy using laser-only AO.
\par Off-axis guiding is less applicable to AO systems on intermediate-class telescopes, such as on the Palomar 60-inch \citep{2014ApJ...790L...8B} or Kitt Peak 2.1-meter \citep{2017RebeccaJC}, the respective past and present host telescopes to Robo-AO. Scintillation also impacts the performance of laser-only AO on smaller apertures more than on larger telescopes, but the loss in effective seeing has been poorly characterized for large numbers of targets on smaller telescopes.  
\par
\begin{figure}[H]
	\centering
	\subfigure[binary star]
	{
		\includegraphics[trim= 10 55 10 100, clip, width=3.4in]{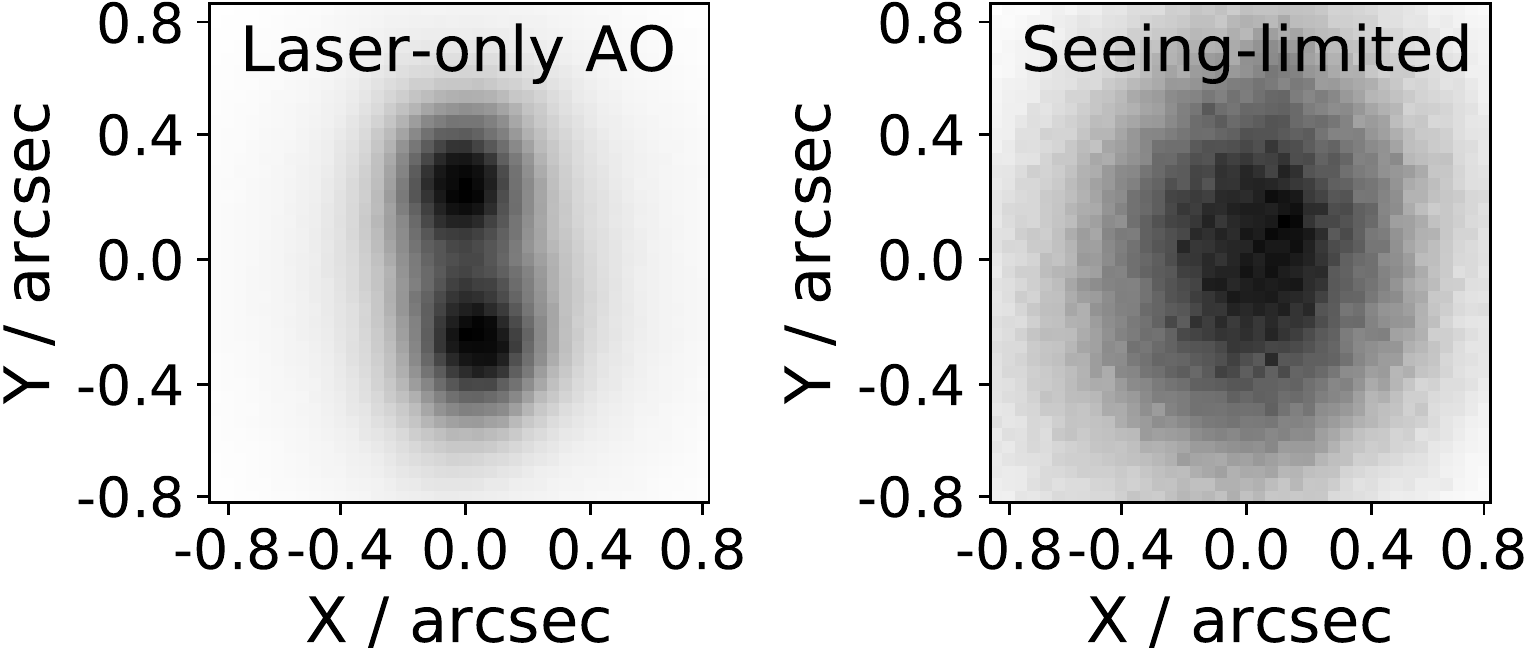}
		\label{fig:compareBIN}
	}
	\subfigure[single star]
	{
		\includegraphics[trim= 10 55 10 100, clip, width=3.4in]{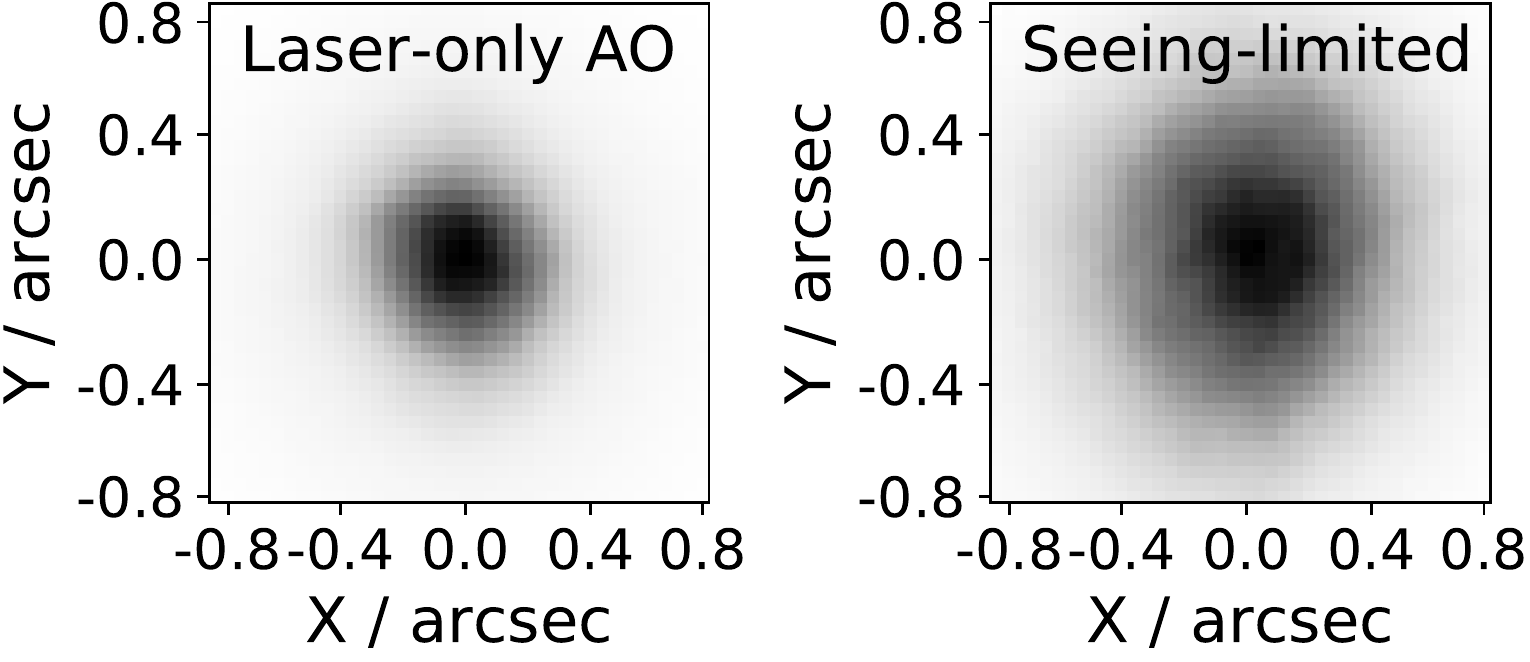}
		\label{fig:compareSIN}
	}
	\caption{No tip-tilt + laser observations from GenSTAC, typifying improvement above the seeing-limit for a binary star \ref{fig:compareBIN} and a single star \ref{fig:compareSIN}. Both images are selected from the one-second (no-tip-tilt correction), FWHM median-improvement bin of Figure \ref{fig:histIMPRV} and displayed with min-max scaling.}
	\label{fig:binsin_compare}
\end{figure}

 We evaluate the effectiveness of laser-only AO systems without tip-tilt correction as an approach to sky coverage limitations for faint targets, but do so for $10^4$ targets, using a vastly smaller robotic LGS system than the VLT or Gemini-North, and in the visible instead of NIR. To do this, we employ a new observation pipeline, \textbf{Gen}eralized \textbf{S}tellar \textbf{T}racking \textbf{A}nd \textbf{C}orrection (GenSTAC), described in section \ref{Laser-only Pipeline} of this paper. Robo-AO+GenSTAC (see Figure \ref{fig:binsin_compare}) reaches targets from the Robo-AO guide star limiting-magnitude of $\sim m_V=15.5$ to the telescope limiting magnitude for a given exposure time. GenSTAC achieves sky coverage at visible wavelengths through its ability to point anywhere in the sky, but at the cost of reduced angular resolution. For arbitrarily-long exposures, GenSTAC would need to be replaced with a physical autoguider to best perform laser-only correction.
\par GenSTAC converts stars that are too faint for guide star correction into sufficiently bright stars through stacking of an integer number of $N$ binned frames and cubic-spline interpolation of stellar drift between averaged stellar positions. Any tip-tilt information inherent in brighter targets is applied for increased improvement to seeing, up to the diffraction limit, much as the current Robo-AO faint-star (i.e. $15.5 < m_V < 18$) pipeline at Kitt Peak does \citep{2017RebeccaJC}. When GenSTAC is instead applied to truly photon-starved targets with no remaining tip-tilt information, laser-only AO provides constant improvement down to the telescope limiting magnitude for a given exposure time.
\par Although GenSTAC's primary purpose is to simulate an autoguider in order to test laser-only AO, it becomes a tip-tilt optimizer when the number of averaged frames is close to one. In this limit, other image registration methods designed for the background-dominated regime of at least a few photons per frame, e.g. \citep{gratadour2005, guillame1998,snyder1990}, may outperform GenSTAC when applied to targets bright enough for partial tip-tilt correction.
\par In Section \ref{Methods}, we describe our observational setup and GenSTAC pipeline design. We also describe our empirical measurements of laser correction with the Robo-AO database. In Section \ref{results}, we describe the results of our measurement of the relative contributions of tip-tilt and laser correction; we also describe the performance of GenSTAC from targets bright enough for full tip-tilt correction to laser-only targets at the limiting magnitude of the telescope for a given exposure time. In Section \ref{conclusion} we conclude and provide recommendations for the future use of laser-only AO.  

\section{Methods}\label{Methods}
To measure the performance of operating in a laser-only mode, we first reduce observations from the Robo-AO system with our laser-only pipeline:
\subsection{Observations and Instrument Setup}
\par Observations were acquired from May of 2012 through June of 2015 with Robo-AO, mounted on the Palomar 60-inch (1.5 m) telescope. Robo-AO efficiently observes hundreds of targets in a night \citep{2014ApJ...790L...8B}. As such, the observed-targets database resulting from hundreds of nights over four years of Robo-AO operation allows quantification of AO performance on a large scale.
\par While the AO system sets up for a given target, it obtains a 20-second seeing-limited observation to assess seeing conditions. The laser-launch system then pulses a 
\begin{figure}[H]
	\centering
	{
		\includegraphics[trim=25 15 18 15, clip, width=3.2in]{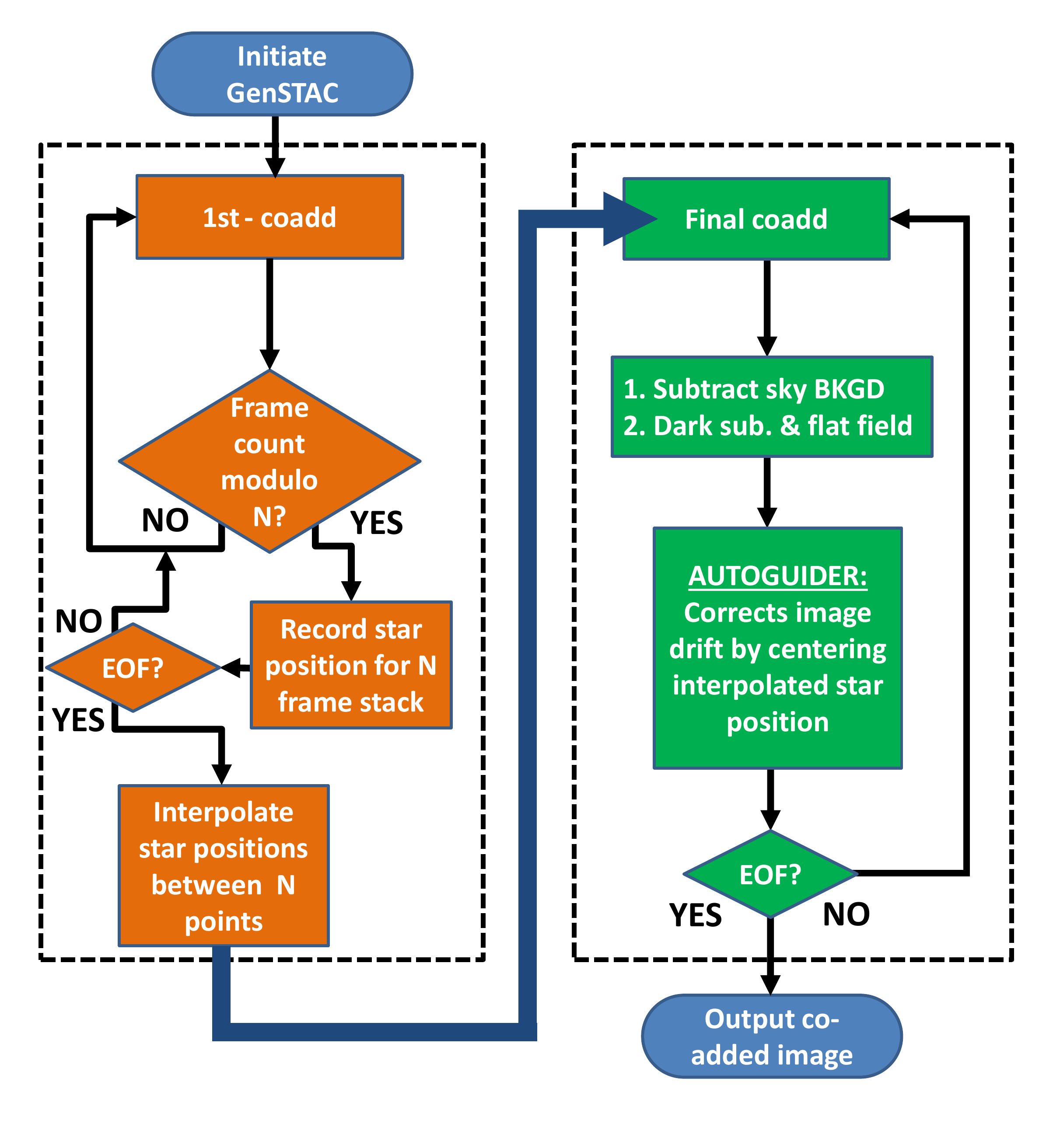}
	}
	\caption{A flowchart describing GenSTAC. The first coaddition step (orange, left) reads in data and averages stellar position over $N$ frames for interpolation. Next, the actual align-and-stack step (green, right) outputs final images. Initiation and output (blue) are also included. ``EOF" abbreviates ``End-of-file."  In this work, we choose $N$ during the GenSTAC initiation step, but this choice may be scripted for fully automated use.}
	\label{fig:pipeline}
\end{figure}
12 W, 355 nm ultraviolet beam along the host telescope line-of-sight out to a distance of 10 km to obtain a measurement of the atmospheric turbulence wavefront. A Shack-Hartmann wavefront sensor in the adaptive optics 1.2 kHz control loop feeds information to a CPU driving a MEMS actuator system, which adapts the shape of a deformable mirror in the optical path, providing wavefront correction. An \textit{Andor iXon} EMCCD science camera records images at 8.6 FPS (0.1168 second frametime), allowing sufficiently-short frametimes for after-the-fact guide star correction of tip-tilt errors, in a reduction pipeline described in \citet{0004-637X-791-1-35}. The EMCCD reduces the read noise for short frametimes from about 50$e^{-}$ to $<$ 1$e^{-}$; Robo-AO employs typical EM-gains between 25x and 300x. The raw data are stored as frames of 1024$\mathrm{\times}$1024 pixels in FITS datacubes with a pixel scale of 0 \farcs 043.

\subsection{Generalized Tracking and Correction Pipeline}\label{Laser-only Pipeline}
We introduce a new pipeline, GenSTAC, to handle laser-only targets. The standard Robo-AO pipeline shifts-and-adds based upon cross-correlation to a diffraction-limited PSF. GenSTAC operates under the assumption of a large Gaussian PSF due to the lack of tip-tilt correction, and hence shifts-and-adds each frame based on averaged stellar positions over many frames. This enables laser-only correction on faint targets and faster operation on bright targets. When the average stellar position over N frames is reduced to $N=1$, the original pipeline outperforms GenSTAC because GenSTAC does not cross-correlate with a diffraction-limited PSF. When GenSTAC finds the best solution to be $N=1$ as described below (i.e. for $m_V < 15.5$ targets), it reverts to the original pipeline for full tip-tilt correction.
\par GenSTAC shifts-and-adds frames according to the following steps, summarized in figure \ref{fig:pipeline}.
\begin{enumerate}
\item \textbf{Raw frames} are read into GenSTAC.
\item \textbf{Choose the number of frames $N$ to bin} together for averaged stellar positions. When scripted, multiple values of $N$ are tested and the best resulting FWHM determines the best$^2$ $N$. \footnotetext[2]{Using FWHM to determine the best $N$ rather than magnitude avoids any danger from incorrect catalog magnitudes.} Otherwise, the user chooses $N$ directly.
\par GenSTAC binning is usually performed based upon the brightness of the target star. Extremely faint objects, producing less than one photon per frame, are binned on second-or-longer timescales to remove long-term tracking drifts. Brighter objects enable faster operation with improved performance using some tip/tilt information. 
\par To quantify the effective seeing of laser-only AO for 15,000 targets, however, $N$ is specified by the user in order to bin away all tip-tilt information regardless of target brightness.
\item A \textbf{first-pass through the frames} is performed. The average position of the star for each group of $N$ frames throughout the observation is recorded during this first coaddition step.
\item The \textbf{averaged pixel positions are interpolated} with one-dimensional cubic splines.
\item The \textbf{frames are dark-subtracted, flat-fielded, shifted, and added} to produce a final image. Shifting uses the output of the cubic interpolation in the previous step. While the standard Robo-AO bright-star pipeline \citep{0004-637X-791-1-35} employs the Drizzle algorithm \citep{drizzleF}, Drizzle is not applied by GenSTAC due to the assumption of large Gaussian PSFs.
\end{enumerate}

\subsection{Measuring Performance} 
Once all frames are reduced into a final image, we use three measures to characterize the quality of each observation: 
the full-width at half-maximum (FWHM) diameter, the 50\% encircled energy ($\theta_{50}$) diameter, and the Strehl ratio. FWHM is calculated as the diameter of the circle with an area of all pixels in the photometric aperture brighter than half the peak. $\theta_{50}$ is calculated as the diameter of the area containing half of the cumulative flux from the PSF. The Strehl is calculated as the total-flux-normalized peak intensity of the PSF divided by a theoretical diffraction-limited, normalized PSF. For the faint targets of the current paper, we are in the very low Strehl regime, making the FWHM and $\theta_{50}$ more relevant in quantifying faint AO performance.
\par We break up 15,000 separately-targeted Robo-AO observations into 42,000 images, typically of 30 seconds 
\begin{figure}[H]
	\centering
	\subfigure[FWHM Comparison]
	{
		\includegraphics[trim= 300 10 40 35,clip, width=3.3in]{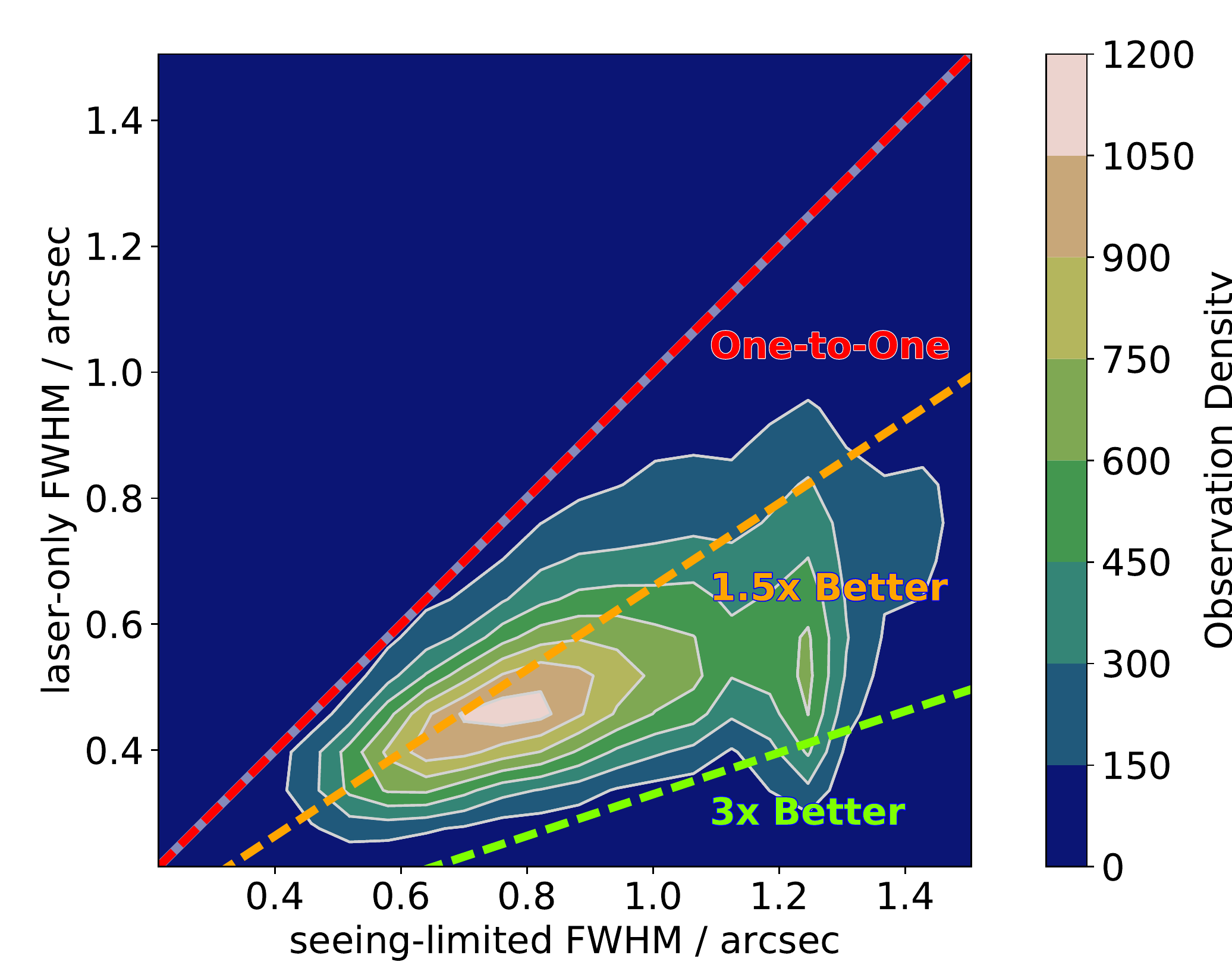}
		\label{fig:compareFWHM}
	}
	\subfigure[$\theta_{50}$ Comparison]
	{
		\includegraphics[trim= 300 10 40 35, clip, width=3.3in]{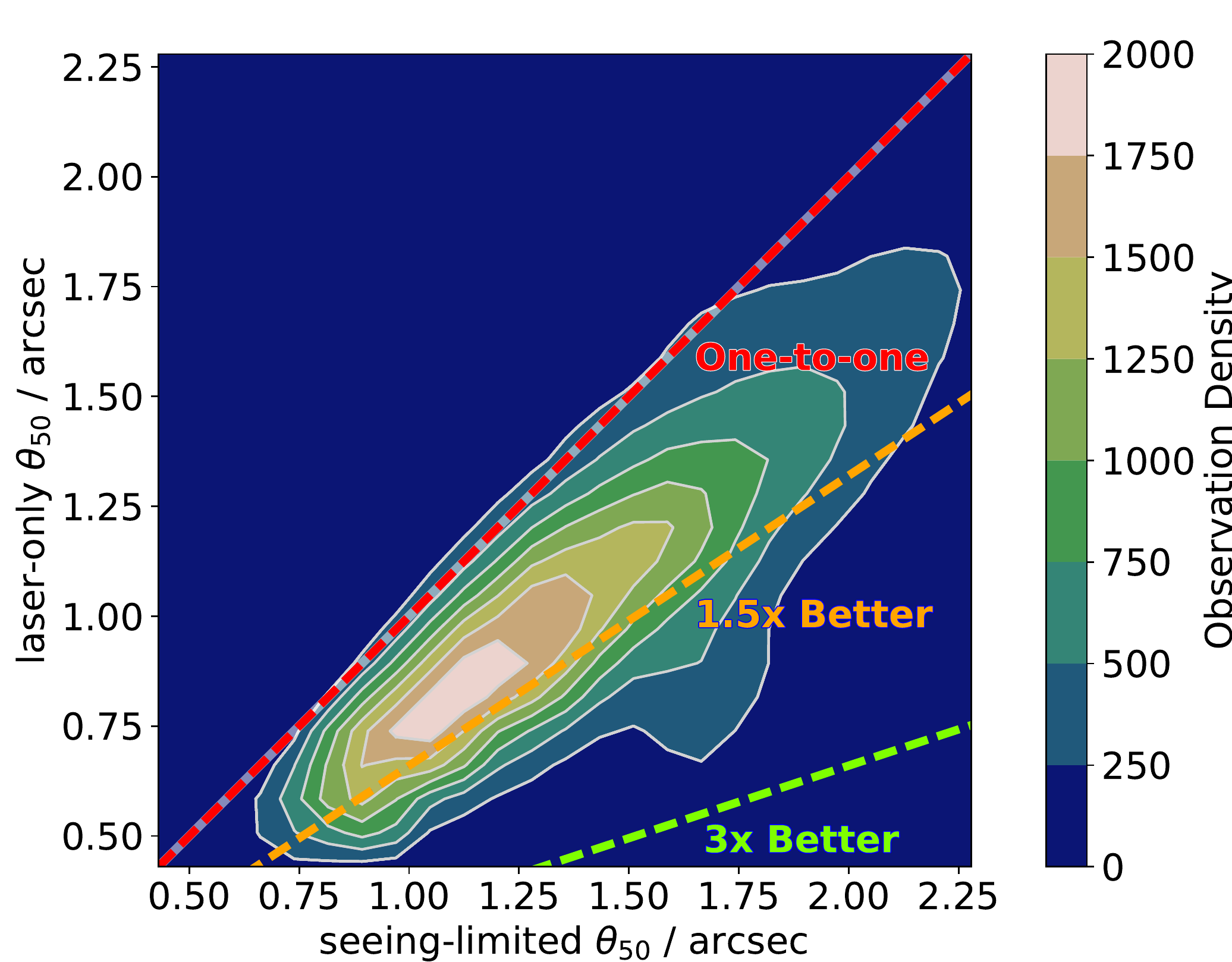}
		\label{fig:compareFWHEF}
	}
	\caption{Performance contour plots of laser-only AO (Y/arcsec) versus seeing-limited performance (X/arcsec). 42,000 raw scatterplot points were 2D-binned, interpolated, and contour-plotted. Reference lines of increasing $\mathrm{FWHM}_\mathrm{seeing}/ \mathrm{FWHM}_\mathrm{laser-only}$ are drawn to guide the eye. These include no increase (red) 1.5x (orange), and 3x (magenta).}
	\label{fig:performance_compare}
\end{figure}
each. We desire to test laser-only image-quality over a wide range of conditions; as the seeing evolves rapidly, these 30-second exposures represent independent realizations of the seeing. We then characterize the FWHM-, $\theta_{50}$-, and Strehl- improvements in effective seeing. To ensure breaking up the 15,000 observations in this way does not bias the seeing statistics, we compared median FWHM and $\theta_{50}$ using only one 30-second image per separately-targeted observation and observed no change in seeing statistics from using 42,000 realizations.

\section{Results}\label{results}
We present measurements over the 15,000-observation Robo-AO dataset of 42,000 independent realizations of the seeing, to show effective seeing for telescope-magnitude-limited targets. Although almost all Robo-AO observed targets were bright enough for tip-tilt correction (by design), we exclude all tip-tilt information from bright targets by binning frame-by-frame changes 
\begin{figure}[H]
	\centering
	\subfigure[FWHM Improvement versus Magnitude]
	{
		\includegraphics[trim= 300 10 40 32,clip, width=3.3in]{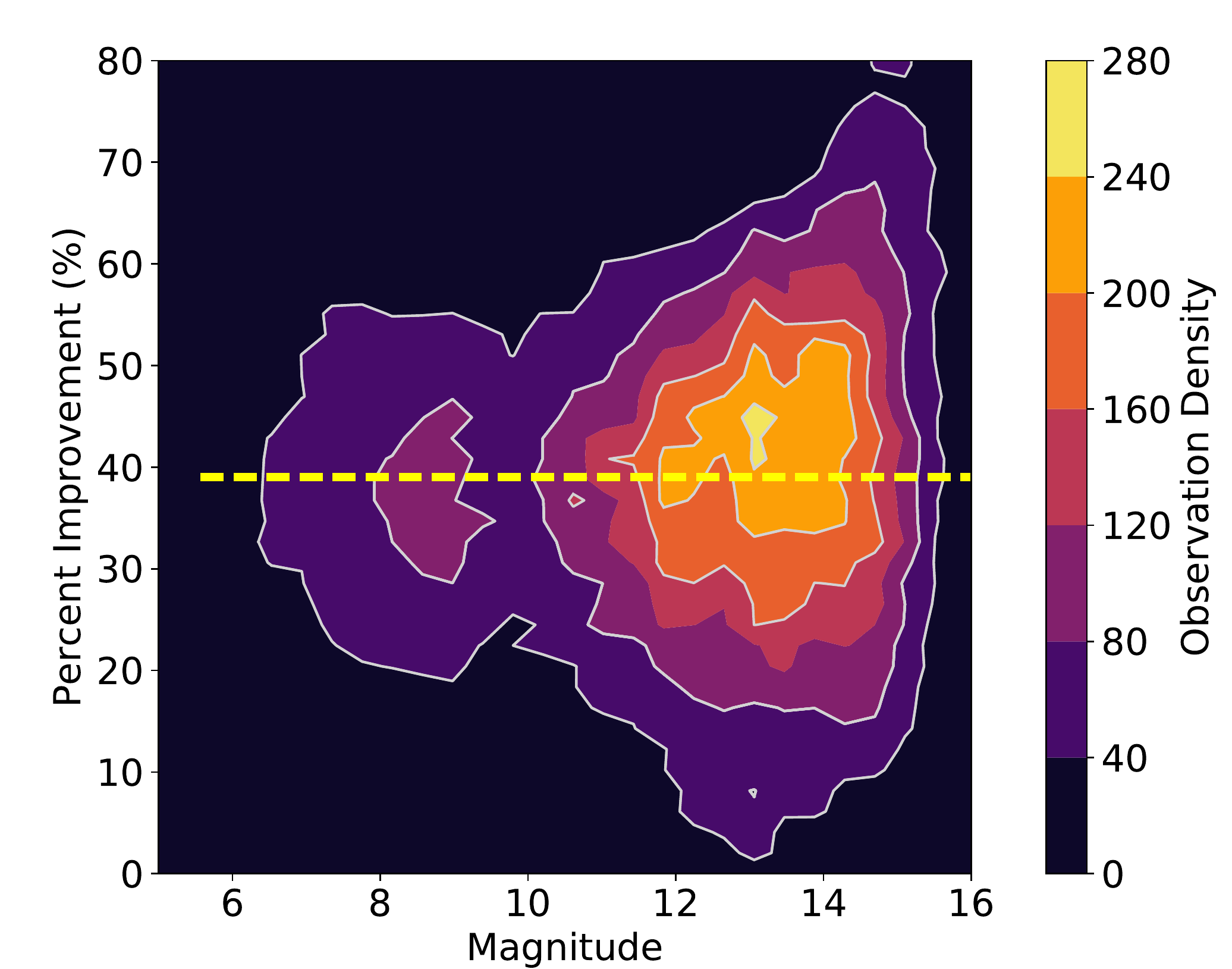}
		\label{fig:imprvFWHM}
	}
	\subfigure[$\theta_{50}$ Improvement versus Magnitude]
	{
		\includegraphics[trim= 300 10 40 32,clip, width=3.3in]{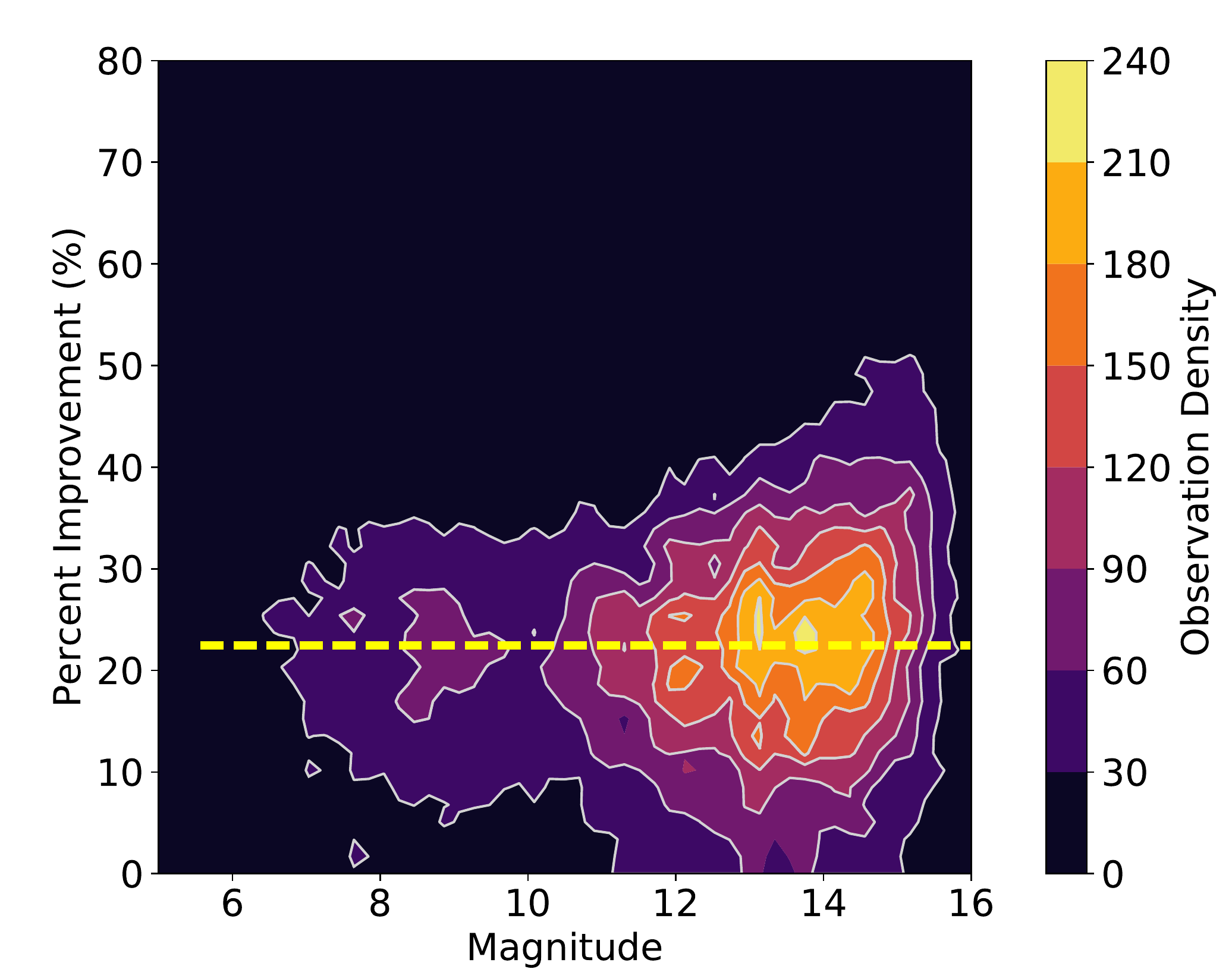}
		\label{fig:imprvFWHEF}
	}
	\caption{Laser-only Improvement versus magnitude. Though median improvement (yellow) remains constant across magnitude, the spread of the data increases toward the faint end due to the effects of noise while measuring resolution.}
	\label{fig:imprvTargets}
\end{figure}
in position away in order to test laser-only AO performance on 15,000 targets. We also run statistical samples of the dataset over a range of binning timescales to quantify effective seeing as tip-tilt error is added, from guide star correction all the way to laser-only correction.

\subsection{System Performance and Resolution Improvement with Laser-only Correction}\label{Performance}
For a 1.2-second binning timescale, simulating an autoguider-equipped Robo-AO for observing a faint star, we compute the seeing-limited and laser-only FWHM, $\theta_{50}$, and Strehl of 14,954 separately-targeted observations, with 42,752 30-second images processed by GenSTAC. Seeing-limited performance is measured from the 20-second acquisition image for each observation. Observations with technical problems that would otherwise bias our data were removed from this list with $5 \sigma$-clips, leaving 14,158 targets and 40,521 images. The resulting distributions, descibed by median value and standard deviation, are as follows: 
\begin{table}
\renewcommand{\arraystretch}{1.6}
\caption{Laser-only AO Results Summary}
\begin{tabular}{ p{1.9cm}|p{1.4cm} p{1.4cm} p{1.4cm} p{1.1cm} }
 \hline
 \textbf{ } & \textbf{Laser-only \newline (arcsec)} & \textbf{Seeing-limited \newline (arcsec)} & \textbf{Improve-ment (\%)} & \textbf{Laser/\newline Seeing \newline Ratio} \\
 \hline
 FWHM   &  0.6$\pm$0.3  & 1.1$\pm$0.5 & 39$\pm$19\% & 0.6$\pm$0.2 \\
 $\theta_{50}$ (50\% enc.)&  1.1$\pm$0.5  & 1.5$\pm$0.5 & 23$\pm$16\% & 0.8$\pm$0.2\\
 Strehl & 0.024$\pm$0.017 & 0.010$\pm$0.007 & ~~~--- & 2.1$\pm$1.1 \\
 \hline
\end{tabular}
{\newline\newline \textbf{Notes}.~-- The Strehl laser/seeing ratio is equivalent to the flux-normalized PSF relative peak intensity and is computed by dividing the laser-only Strehl by the seeing-limited Strehl for each observation. A $5\sigma$ cut was applied to the Strehls due to a small number of biasing outliers.}
\newcommand\tab[1][1cm]{\hspace*{#1}}
{\newline \tab[0.98cm] -- Laser-only values, seeing-limited values, percent-improvements, and ratios are computed for each observation separately. Medians and standard errors of the distributions are given; hence computing improvements directly from the quoted measurements will produce values different from the ones above.}
\renewcommand{\arraystretch}{10}
\end{table}
\hfill \break
Seeing-limited FWHM is 1\farcs 1$\pm$0\farcs 5, while laser-only AO FWHM is 0\farcs 6$\pm$0\farcs 3, an improvement of 39$\pm$19\%. Seeing-limited $\theta_{50}$ is 1\farcs 5$\pm$0\farcs 5, while laser-only AO $\theta_{50}$ is 1\farcs 1$\pm$0\farcs 5, an improvement of 23$\pm$16\%. Strehls also showed improvement, although Strehl contrast in the faint-target regime becomes less relevant due to lower values. Laser-only Strehl is 0.024$\pm$0.017, while the seeing-limited Strehl is 0.010$\pm$0.007. Strehl improvement over the seeing-limit and the associated uncertainty are large due to the non-linear peak effects of concentrating photons into a smaller area. All 42,000 individual observations are plotted against each other in Figure \ref{fig:performance_compare}. Scatterplot points are 2D-binned, interpolated, and displayed as AO-resolution versus seeing-resolution contours. 
\par As a check against implicit tuning toward bright targets in the Robo-AO dataset, we examine improvement-vs-brightness contour-scatterplots in Figure \ref{fig:imprvTargets}, which remains relatively constant across magnitude. For faint targets close to the background-noise level, measurement of resolution becomes slightly less accurate, with a small apparent increase in improvement, although not at a significant-enough level to obscure the constant-improvement trend.
\begin{figure*}
	\setcounter{figure}{5}
	\centering
	\subfigure[Laser-off / Laser-on FWHM Distributions]
	{
		\includegraphics[trim= 5 1 10 1,clip, width=6.6in]{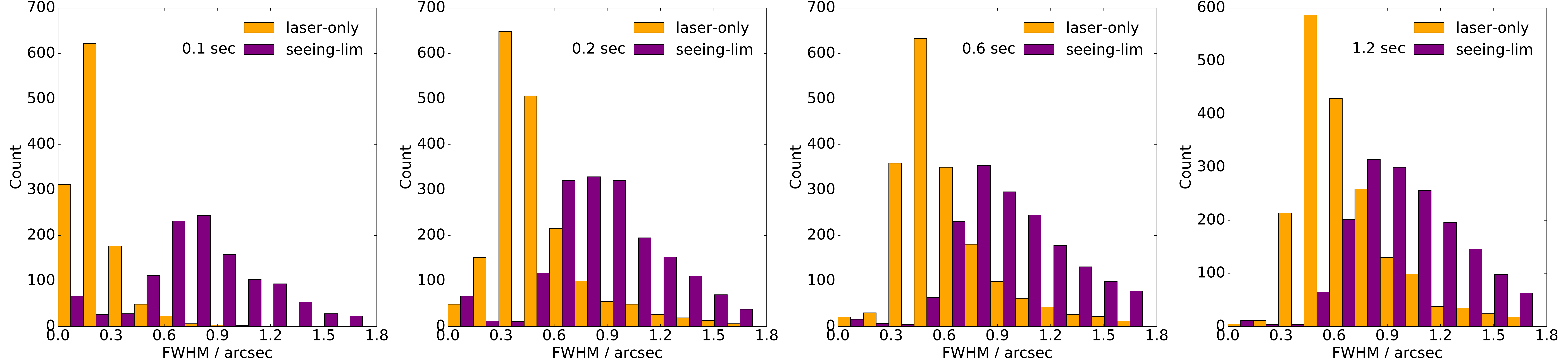}
		\label{fig:histFWHM}
	}
	\subfigure[Laser-off / Laser-on $\theta_{50}$ Distributions]
	{
		\includegraphics[trim= 7 1 10 1, clip, width=6.6in]{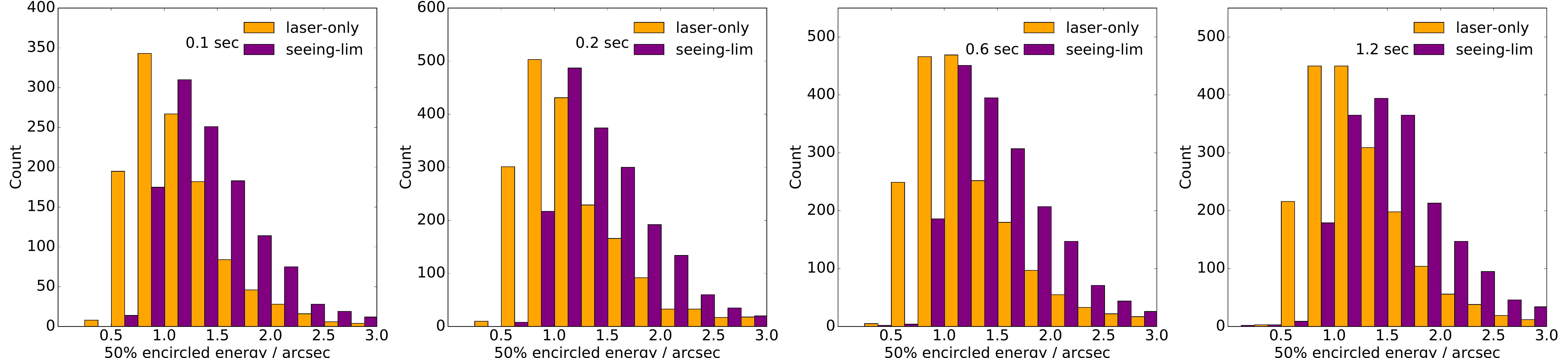}
		\label{fig:histFWHEF}
	}
	\subfigure[FWHM and $\theta_{50}$ Improvement Distributions]
	{
		\includegraphics[trim= 5 1 10 1, clip, width=6.6in]{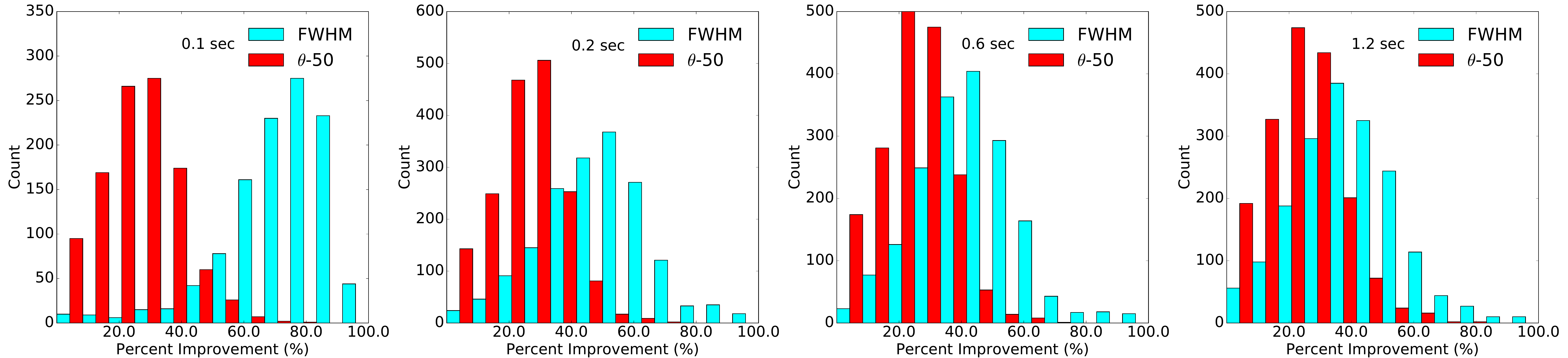}
		\label{fig:histIMPRV}
	}
	\caption{Well-behaved normal or skewed-normal resolution distributions, for laser-on and laser-off. Distribution medians indicate system performance, as distribution wings contain some observations with technical problems. However much tip-tilt is possible for each timescale is applied to both seeing-limited and AO-images, resulting in slightly-underestimated improvement for 0.1-0.6 second timescales.}
	\label{fig:hist_compare}
\end{figure*}
\subsection{Sensitivity of Improvement to Guiding Timescale}\label{auto_variation}
\par In addition to laser-only correction, we measure AO-correction with varying amounts of tip-tilt present. Stacking more frames progressively removes tip-tilt information. In Figure \ref{fig:auto_vars} we show that the FWHM performance drops to a steady value around one-second timescales, while the $\theta_{50}$ remains approximately constant across timescales. 
\par We randomly selected several observations of $\theta_{50}$ median improvement and measured normalized cumulative flux as a function of radius away from PSF center. For each target, the cumulative flux distribution for full-AO, partial tip-tilt ``hybrid"-AO, and seeing-limited tracking were compared. A 0.6-second guiding timescale was chosen as representative of ``hybrid" tracking. While seeing-limited cumulative flux always converges last, both full-AO and ``hybrid"-AO distributions quickly converge and have nearly identical $\theta_{50}$, within $\sim 0 \farcs 1$ of each other. Within the 50\% encircled energy radius, the contributions from the PSF core to the cumulative distribution, which occur at very small radii, are dominated by the roughly linear increase in PSF flux that scales as enclosed-area. 
This agrees with the median $\theta_{50}$ obtained from each larger guiding-timescale sample: a diffraction-limited 0.1-second timescale median $\theta_{50}$ of 1 \farcs 04 is only 0 \farcs 1 better than a 1.2-second autoguider timescale median $\theta_{50}$ of 1 \farcs 15. For reference, the median $\theta_{50}$ for slow-guided, seeing-limited observations is 1 \farcs 5. 
\par The takeaway from this 50\% encircled energy behavior is that for science goals requiring high-resolution energy-encircled observations, a Robo-AO LGS system equipped with an autoguider and running the laser system should give comparable performance to full-AO correction. For faint targets especially, this greatly reduces the sky-coverage tip-tilt problem. 
\par We also estimate the limiting magnitude for which each timescale is valid. Diffraction-limited performance with tip-tilt correction is possible to {\raise.17ex\hbox{$\scriptstyle\mathtt{‌​\sim}$}}15.5th magnitude. We estimate that GenSTAC provides some improvement by using residual tip-tilt information down to {\raise.17ex\hbox{$\scriptstyle\mathtt{‌​\sim}$}}18th magnitude, and operates as a software slow-guider for targets down to the limiting magnitude of the instrument at a given exposure time.
\par Samples were randomly selected from 42,000 observations and processed at characteristic timescales ($N$). 
\begin{figure}[H]
	\setcounter{figure}{4}
	\centering
	{
		\includegraphics[trim= 70 10 100 10,clip, width=3.4in]{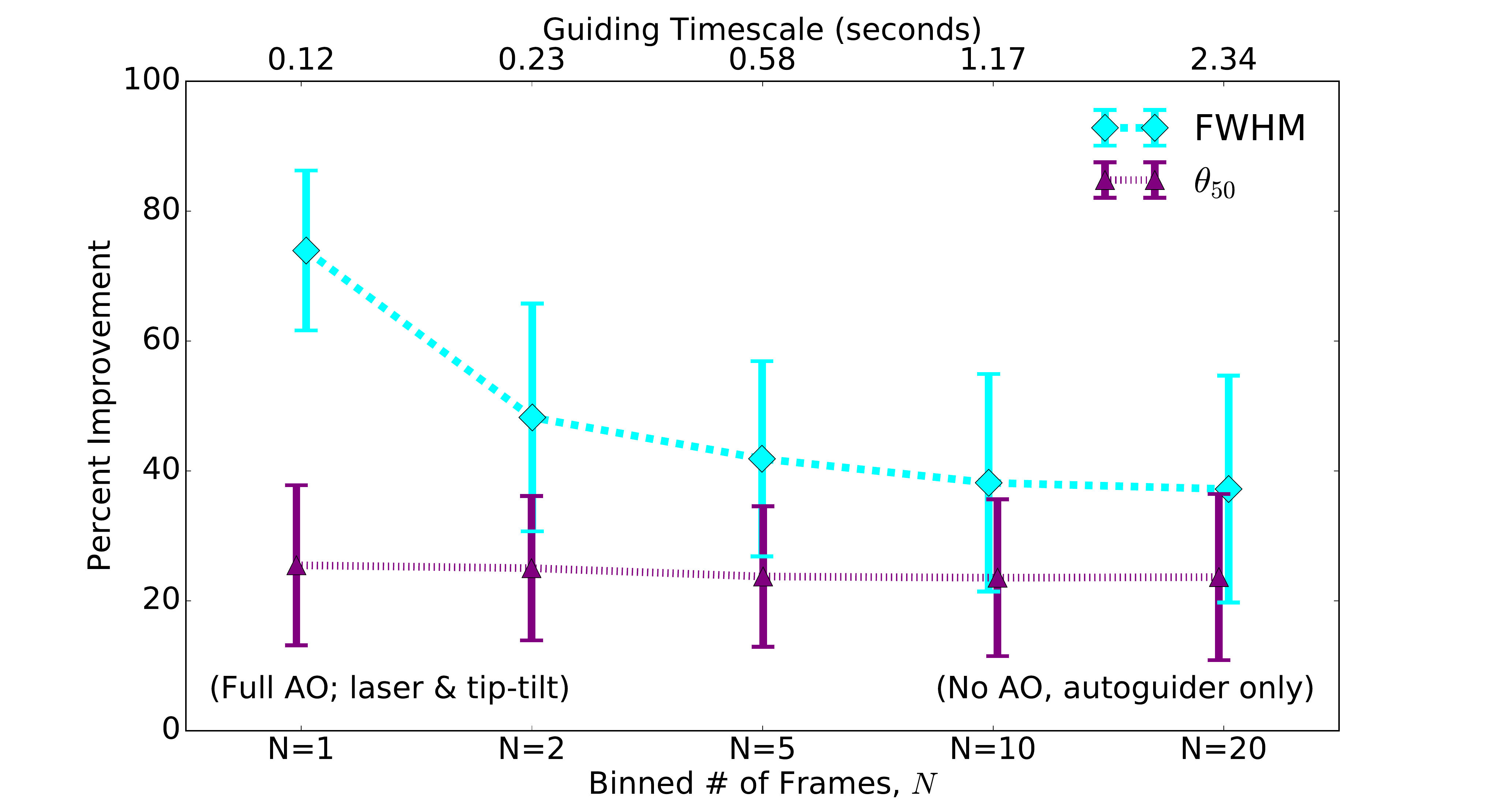}
	}
	\caption{Improvement in effective seeing across guiding timescale. GenSTAC's guide star mode lies to the left, where full tip-tilt information is employed, while the autoguider mode to the right uses no tip-tilt information. $3 \sigma$-clips are applied to the means of each FWHM distribution, as FWHM standard performance in ``hybrid mode" is affected by failed measurements in the distribution wings. $\theta_{50}$ is more robust, requiring no cuts.}
	\label{fig:auto_vars}
\end{figure}
Assuming approximately normal distributions, mean improvements for each timescale were obtained at a 99\% confidence interval and margins-of-error of no more than {\raise.17ex\hbox{$\scriptstyle\mathtt{‌​\sim}$}}1.1 pixel and {\raise.17ex\hbox{$\scriptstyle\mathtt{‌​\sim}$}}2\% for resolutions (FWHM and $\theta_{50}$) and percent improvements, respectively. A $3\sigma$ cut to FWHM was applied to each sample, removing up to 10\% of observations. Without this cut, mean FWHM improvement is affected by occasional observations with technical problems, where telescope shake was present, or the PSF of a binary was measured assuming a non-binary. The margin-of-error for FWHM and $\theta_{50}$ remained less than {\raise.17ex\hbox{$\scriptstyle\mathtt{‌​\sim}$}}1.1 pixel without any sigma-clipping. Figure \ref{fig:hist_compare} verifies our assumption of normal or skewed-normal distributions for each measurement, such that $\sigma$-clipping does not remove standard system performance.

\section{Discussion and Conclusions}\label{conclusion}
\par For our LGS AO system, Robo-AO, the effective seeing upon arbitrarily faint targets demonstrates 39$\pm$19\% improvement above the seeing-limit, even in the absence of tip-tilt correction. Faint targets with some recoverable tip-tilt information show corresponding gains in seeing, culminating in diffraction-limited performance for moderately bright guide star targets. Our FWHM results on an intermediate-class telescope show a 2x-3x effective-seeing improvement similar to those of the LGS+P1 mode upon the Gemini-North telescope. Due to the lack of tip-tilt correction, laser-only AO does not require large apertures such as that of Gemini-North. Furthermore, whereas LGS+P1 improvements have been demonstrated in the NIR for $10^2$ targets \citep{2013aoel.confE..51T}, our LGS without tip-tilt improvement holds for $10^4$ targets, into the visible. 
\par 50\% encircled energy demonstrates for $10^4$ targets the same improvement upon both diffraction-limited tip-tilt correction and slow-guiding correction where tip-tilt correction is no longer possible, in agreement with \citet{2008Msngr.131....7D}. These results suggest that intermediate-class telescopes equipped with LGS AO systems will see significant gains in performance upon faint targets when running the laser with a slow-guiding drift corrector, especially upon targets for which we most care about the energy encircled. For example, the proposed Rapid Transient Surveyor instrument (RTS) combines an integral field spectrograph (IFS) placed behind an LGS AO system on the robotized University of Hawai'i 2.2-meter telescope. RTS will enable precise mapping of dark matter in the local universe by characterizing SN1a in the IR, which requires IFS for faint sources \citep{2016SPIE.9909E..0FB}. Our $\theta_{50}$ results motivate future exploration of laser-only IFS for those RTS targets without sufficiently-bright guide stars.
\par We investigate laser-only AO with GenSTAC, a new Robo-AO pipeline. GenSTAC provides post-processing options for guide-star correction, partial tip-tilt correction, and laser-only correction. GenSTAC binning on timescales longer than 1 second serves as a software equivalent to a slow-guider mounted to the telescope by converting stars that are too faint for guide star correction into adequate stars through averaging and interpolating procedures described in \ref{Laser-only Pipeline}. For moderately-faint targets, it extracts any remaining tip-tilt information; very faint targets with no tip-tilt information will rely purely on laser correction.
\par In future work, we employ laser-only correction to search hundreds of faint KOIs ($m_\textit{Kepler}>15.7$) for stellar blends with Robo-AO and compare the companion fraction between the faint and bright KOI populations, especially for M-dwarfs.  
\par As LGS AO systems continue to proliferate \citep{2012ARA&A..50..305D}, small-to-intermediate-class observatories with LGS will benefit greatly from running the laser and a slow-guider whenever observing faint targets beyond diffraction-limited sky-coverage.

\section{Acknowledgements}\label{discussion}
We thank Rebecca Jensen-Clem for helpful discussions on the project. 
\par This research has made use of the NASA Exoplanet Archive, which is operated by the California Institute of Technology, under contract with the National Aeronautics and Space Administration under the Exoplanet Exploration Program.
\par This research made use of Astropy, a community-developed core Python package for Astronomy \citep{2013A&A...558A..33A}, and the NumPy, SciPy, and Matplotlib Python modules \citep{numpyscipy,Jones2001,matplotlib}.
\par This work was partially supported by NASA XRP Grant NNX15AC91G. The Robo-AO system is supported by collaborating partner institutions, the California Institute of Technology and the Inter-University Centre for Astronomy and Astrophysics, and by the National Science Foundation under Grant Nos. AST-0906060,  AST-0960343, and AST-1207891, by the Mount Cuba Astronomical Foundation, and by a gift from Samuel Oschin. C.B. acknowledges support from the Alfred P. Sloan Foundation. We are grateful to the Palomar Observatory staff for their support of Robo-AO on the 1.5-m telescope, particularly S. Kunsman, M. Doyle, J. Henning, R. Walters, G. Van Idsinga, B. Baker, K. Dunscombe and D. Roderick.
\par This research was carried out on the traditional lands of the Occaneechi, Haw, and Eno Native American tribes in Orange County, NC, and on the traditional lands of the Luise\~no, or Pay\'omkawichum, at Palomar Observatory, CA.
\par Finally, we would like to thank the anonymous referee who graciously gave their time to make this the best version of this work.

\bibliographystyle{apj}
\bibliography{paper_references}

\begin{thebibliography}{}
\expandafter\ifx\csname natexlab\endcsname\relax\def\natexlab#1{#1}\fi

\bibitem[{{Astropy Collaboration} {et~al.}(2013){Astropy Collaboration},
  {Robitaille}, {Tollerud}, {Greenfield}, {Droettboom}, {Bray}, {Aldcroft},
  {Davis}, {Ginsburg}, {Price-Whelan}, {Kerzendorf}, {Conley}, {Crighton},
  {Barbary}, {Muna}, {Ferguson}, {Grollier}, {Parikh}, {Nair}, {Unther},
  {Deil}, {Woillez}, {Conseil}, {Kramer}, {Turner}, {Singer}, {Fox}, {Weaver},
  {Zabalza}, {Edwards}, {Azalee Bostroem}, {Burke}, {Casey}, {Crawford},
  {Dencheva}, {Ely}, {Jenness}, {Labrie}, {Lim}, {Pierfederici}, {Pontzen},
  {Ptak}, {Refsdal}, {Servillat}, \& {Streicher}}]{2013A&A...558A..33A}
{Astropy Collaboration}, {Robitaille}, T.~P., {Tollerud}, E.~J., {et~al.} 2013,
  \aap, 558, A33

\bibitem[{{Baranec} {et~al.}(2016{\natexlab{a}}){Baranec}, {Ziegler}, {Law},
  {Morton}, {Riddle}, {Atkinson}, {Schonhut}, \& {Crepp}}]{2016AJ....152...18B}
{Baranec}, C., {Ziegler}, C., {Law}, N.~M., {et~al.} 2016{\natexlab{a}}, \aj,
  152, 18

\bibitem[{{Baranec} {et~al.}(2014){Baranec}, {Riddle}, {Law}, {Ramaprakash},
  {Tendulkar}, {Hogstrom}, {Bui}, {Burse}, {Chordia}, {Das}, {Dekany},
  {Kulkarni}, \& {Punnadi}}]{2014ApJ...790L...8B}
{Baranec}, C., {Riddle}, R., {Law}, N.~M., {et~al.} 2014, \apjl, 790, L8

\bibitem[{{Baranec} {et~al.}(2016{\natexlab{b}}){Baranec}, {Lu}, {Wright},
  {Tonry}, {Tully}, {Szapudi}, {Takamiya}, {Hunter}, {Riddle}, {Chen}, \&
  {Chun}}]{2016SPIE.9909E..0FB}
{Baranec}, C., {Lu}, J.~R., {Wright}, S.~A., {et~al.} 2016{\natexlab{b}}, in
  \procspie, Vol. 9909, Adaptive Optics Systems V, 99090F

\bibitem[{{Chabrier}(2003)}]{2003PASP..115..763C}
{Chabrier}, G. 2003, \pasp, 115, 763

\bibitem[{{Davies} \& {Kasper}(2012)}]{2012ARA&A..50..305D}
{Davies}, R., \& {Kasper}, M. 2012, \araa, 50, 305

\bibitem[{{Davies} {et~al.}(2008){Davies}, {Rabien}, {Lidman}, {Le Louarn},
  {Kasper}, {F{\"o}rster Schreiber}, {Roccatagliata}, {Ageorges}, {Amico},
  {Dumas}, \& {Mannucci}}]{2008Msngr.131....7D}
{Davies}, R., {Rabien}, S., {Lidman}, C., {et~al.} 2008, The Messenger, 131, 7

\bibitem[{{Dressing} \& {Charbonneau}(2015)}]{DressingCharbonneau}
{Dressing}, C.~D., \& {Charbonneau}, D. 2015, \apj, 807, 45

\bibitem[{{Foy} \& {Labeyrie}(1985)}]{1985A&A...152L..29F}
{Foy}, R., \& {Labeyrie}, A. 1985, \aap, 152, L29

\bibitem[{Fruchter \& Hook(2002)}]{drizzleF}
Fruchter, A.~S., \& Hook, R.~N. 2002, Publications of the Astronomical Society
  of the Pacific, 114, 144

\bibitem[{{Gratadour} {et~al.}(2005){Gratadour}, {Mugnier}, \&
  {Rouan}}]{gratadour2005}
{Gratadour}, D., {Mugnier}, L.~M., \& {Rouan}, D. 2005, \aap, 443, 357

\bibitem[{{Guillaume} {et~al.}(1998){Guillaume}, {Melon}, {Refregier}, \&
  {Llebaria}}]{guillame1998}
{Guillaume}, M., {Melon}, P., {Refregier}, P., \& {Llebaria}, A. 1998, Journal
  of the Optical Society of America A, 15, 2841

\bibitem[{Haas {et~al.}(2010)Haas, Batalha, Bryson, Caldwell, Dotson, Hall,
  Jenkins, Klaus, Koch, Kolodziejczak, Middour, Smith, Sobeck, Stober,
  Thompson, \& Cleve}]{2041-8205-713-2-L115}
Haas, M.~R., Batalha, N.~M., Bryson, S.~T., {et~al.} 2010, The Astrophysical
  Journal Letters, 713, L115

\bibitem[{{Hardy}(1998)}]{Hardy1998}
{Hardy}, J.~W. 1998, {Adaptive Optics for Astronomical Telescopes}, 448

\bibitem[{{Hunter}(2007)}]{matplotlib}
{Hunter}, J.~D. 2007, Computing in Science \& Engineering, 9, 90

\bibitem[{{Jensen-Clem} {et~al.}(2017){Jensen-Clem}, {Duev}, {Riddle},
  {Salama}, {Baranec}, {Law}, {Kulkarni}, \& {Ramprakash}}]{2017RebeccaJC}
{Jensen-Clem}, R., {Duev}, D.~A., {Riddle}, R., {et~al.} 2017, ArXiv e-prints,
  arXiv:1703.08867

\bibitem[{Johnson {et~al.}(2011)Johnson, Apps, Gazak, Crepp, Crossfield,
  Howard, Marcy, Morton, Chubak, \& Isaacson}]{0004-637X-730-2-79}
Johnson, J.~A., Apps, K., Gazak, J.~Z., {et~al.} 2011, The Astrophysical
  Journal, 730, 79

\bibitem[{Jones {et~al.}(2001--)Jones, Oliphant, Peterson,
  {et~al.}}]{Jones2001}
Jones, E., Oliphant, T., Peterson, P., {et~al.} 2001--, {SciPy}: Open source
  scientific tools for {Python}

\bibitem[{{Koch} {et~al.}(2010){Koch}, {Borucki}, {Basri}, {Batalha}, {Brown},
  {Caldwell}, {Christensen-Dalsgaard}, {Cochran}, {DeVore}, {Dunham},
  {Gautier}, {Geary}, {Gilliland}, {Gould}, {Jenkins}, {Kondo}, {Latham},
  {Lissauer}, {Marcy}, {Monet}, {Sasselov}, {Boss}, {Brownlee}, {Caldwell},
  {Dupree}, {Howell}, {Kjeldsen}, {Meibom}, {Morrison}, {Owen}, {Reitsema},
  {Tarter}, {Bryson}, {Dotson}, {Gazis}, {Haas}, {Kolodziejczak}, {Rowe}, {Van
  Cleve}, {Allen}, {Chandrasekaran}, {Clarke}, {Li}, {Quintana}, {Tenenbaum},
  {Twicken}, \& {Wu}}]{2010ApJ...713L..79K}
{Koch}, D.~G., {Borucki}, W.~J., {Basri}, G., {et~al.} 2010, \apjl, 713, L79

\bibitem[{Law {et~al.}(2014)Law, Morton, Baranec, Riddle, Ravichandran,
  Ziegler, Johnson, Tendulkar, Bui, Burse, Das, Dekany, Kulkarni, Punnadi, \&
  Ramaprakash}]{0004-637X-791-1-35}
Law, N.~M., Morton, T., Baranec, C., {et~al.} 2014, The Astrophysical Journal,
  791, 35

\bibitem[{{O'Donovan} {et~al.}(2006){O'Donovan}, {Charbonneau}, {Torres},
  {Mandushev}, {Dunham}, {Latham}, {Alonso}, {Brown}, {Esquerdo}, {Everett}, \&
  {Creevey}}]{2006ApJ...644.1237O}
{O'Donovan}, F.~T., {Charbonneau}, D., {Torres}, G., {et~al.} 2006, \apj, 644,
  1237

\bibitem[{{Rigaut} \& {Gendron}(1992)}]{1992A&A...261..677R}
{Rigaut}, F., \& {Gendron}, E. 1992, \aap, 261, 677

\bibitem[{{Roberts} \& {Singh}(1998)}]{1998SPIE.3353..611R}
{Roberts}, S., \& {Singh}, G. 1998, in \procspie, Vol. 3353, Adaptive Optical
  System Technologies, ed. D.~{Bonaccini} \& R.~K. {Tyson}, 611--620

\bibitem[{{Snyder} \& {Schulz}(1990)}]{snyder1990}
{Snyder}, D.~L., \& {Schulz}, T.~J. 1990, Journal of the Optical Society of
  America A, 7, 1251

\bibitem[{{Steinbring} {et~al.}(2005){Steinbring}, {Faber}, {Macintosh},
  {Gavel}, \& {Gates}}]{Steinbring}
{Steinbring}, E., {Faber}, S.~M., {Macintosh}, B.~A., {Gavel}, D., \& {Gates},
  E.~L. 2005, \pasp, 117, 847

\bibitem[{{Trujillo} {et~al.}(2013){Trujillo}, {Ball}, {Boccas}, {Cavedoni},
  {Christou}, {Coulson}, {Ebbers}, {Emig}, {Jorgensen}, {Kang}, {Lai},
  {Matulonis}, {McDermid}, {Miller}, {Neichel}, {Oram}, {Rigaut}, {Roth},
  {Schneider}, {Stephens}, {Trancho}, {Walls}, {White}, \& {Gemini Software
  Team}}]{2013aoel.confE..51T}
{Trujillo}, C., {Ball}, J., {Boccas}, M., {et~al.} 2013, in Proceedings of the
  Third AO4ELT Conference, ed. S.~{Esposito} \& L.~{Fini}, 51

\bibitem[{{van der Walt} {et~al.}(2011){van der Walt}, {Colbert}, \&
  {Varoquaux}}]{numpyscipy}
{van der Walt}, S., {Colbert}, S.~C., \& {Varoquaux}, G. 2011, Computing in
  Science \& Engineering, 13, 22

\bibitem[{{Wizinowich} {et~al.}(2014){Wizinowich}, {Smith}, {Biasi}, {Cetre},
  {Dekany}, {Femenia-Castella}, {Fucik}, {Hale}, {Neyman}, {Pescoller},
  {Ragland}, {Stomski}, {Andrighettoni}, {Bartos}, {Bui}, {Cooper}, {Cromer},
  {van Dam}, {Hess}, {James}, {Lyke}, {Rodriguez}, \& {Stalcup}}]{Wizinowich}
{Wizinowich}, P., {Smith}, R., {Biasi}, R., {et~al.} 2014, in \procspie, Vol.
  9148, Adaptive Optics Systems IV, 91482B

\bibitem[{{Ziegler} {et~al.}(2017){Ziegler}, {Law}, {Morton}, {Baranec},
  {Riddle}, {Atkinson}, {Baker}, {Roberts}, \& {Ciardi}}]{2016arXiv160503584Z}
{Ziegler}, C., {Law}, N.~M., {Morton}, T., {et~al.} 2017, \aj, 153, 66

\end{thebibliography}

\end{document}